\documentclass[aps,twocolumn,floatfix,longbibliography,superscriptaddress]{revtex4-2}
\usepackage[T1]{fontenc}
\usepackage{graphics,graphicx,epsfig}
\usepackage{amssymb,color}
\usepackage{epsf,epstopdf}
\usepackage {amsmath}
\usepackage[final,hidelinks]{hyperref}
\usepackage{ragged2e}
\usepackage{siunitx}

\newcommand{\avg}[1]{\langle{#1}\rangle{}}
\newcommand{\tc}[0]{T_{\rm C}}

\begin{document}

\title{Invariant non-equilibrium dynamics of transcriptional regulation optimize information flow}

\author{Benjamin Zoller}
\affiliation{Department of Stem Cell and Developmental Biology, CNRS UMR3738 Paris Cité, Institut Pasteur, 25 rue du Docteur Roux, FR-75015 Paris, France}
\author{Alexis Bénichou}
\affiliation{Institute of Science and Technology Austria, Am Campus 1, AT-3400 Klosterneuburg, Austria}
\author{Thomas Gregor}
\affiliation{Joseph Henry Laboratories of Physics \& Lewis-Sigler Institute for Integrative Genomics, Princeton University, Princeton, NJ 08544, USA}
\affiliation{Department of Stem Cell and Developmental Biology, CNRS UMR3738 Paris Cité, Institut Pasteur, 25 rue du Docteur Roux, FR-75015 Paris, France}
\author{Ga\v{s}per Tka\v{c}ik}
\affiliation{Institute of Science and Technology Austria, Am Campus 1, AT-3400 Klosterneuburg, Austria}
 
\date{\today}
%TC:ignore
\begin{abstract}
% right now right about 200 words, OK for PNAS and Nat Phys
Eukaryotic gene regulation is based on stochastic yet controlled promoter switching, during which genes transition between transcriptionally active and inactive states. Despite the molecular complexity of this process, recent studies reveal a surprising invariance of the ``switching correlation time'' ($\tc$), which characterizes promoter activity fluctuations, across gene expression levels in diverse genes and organisms. A biophysically plausible explanation for this invariance remains missing. 
Here, we show that this invariance imposes stringent constraints on minimal yet plausible models of transcriptional regulation, requiring at least four system states and non-equilibrium dynamics that break detailed balance. Using Bayesian inference on \textit{Drosophila} gap gene expression data, we demonstrate that such models (i) accurately reproduce the observed $\tc$-invariance; (ii) remain robust to parameter perturbations; and (iii) maximize information transmission from transcription factor concentration to gene expression. These findings suggest that eukaryotic gene regulation has evolved to balance precision with reaction rate and energy dissipation constraints, favoring non-equilibrium architectures for optimal information transmission.
\end{abstract}
%TC:endignore

\maketitle

\section*{Introduction}

Transcriptional regulation enables cells to control gene expression in response to environmental and developmental cues. At the single-cell level, this process is inherently stochastic, with transcription occurring in random bursts as promoters switch between inactive and active states~\cite{Rodriguez:2020,Meeussen:2024}. The frequency and duration of these bursts are influenced by transcription factors (TFs), cis-regulatory elements (e.g., enhancers), and chromatin accessibility, shaping gene expression dynamics in ways essential for cellular function but challenging to analyze~\cite{Jonge:2022,Wagh:2023}. Advances in live-cell imaging and single-molecule tracking have provided unprecedented insight into these dynamics, enabling increasingly detailed mathematical models to be inferred from experimental data~\cite{Rodriguez:2018eo,Wan:2021,Tantale:2021,Pimmett:2021,Chen:2023,Trzaskoma:2024}.

Recent precision measurements of gene expression during early \emph{Drosophila} development established that the switching correlation time for promoter activity, $\tc$, remains nearly constant across gene expression levels~\cite{Zoller:2018gj,Chen:2023}. Further analyses of published data suggest that this phenomenon---referred to as $\tc$-invariance---is conserved across multiple genes and organisms~\cite{Chen:2023}. $\tc$ is a key dynamical observable that characterizes the timescale over which promoters fluctuate between ON and OFF states. It sets both the relaxation timescale toward steady state and the window over which gene expression fluctuations can be effectively averaged or buffered \cite{Paulsson:2005da,Raj:2006gq,Tkacik:2008ht}. 

This invariance is unexpected for several reasons. First, transcriptional regulation involves multiple molecular interactions, reaction rates, and external inputs, making the observation that simple ``effective'' two-state models  capture experimental data well extremely surprising \cite{Zoller:2018gj,Chen:2023}. Second, even if such a two-state approximation holds, explaining $\tc$-invariance with respect to gene expression level requires intricate parameter coordination that lacks an obvious mechanistic basis~\cite{Meeussen:2024}. Third, standard biochemical models of transcription do not predict such invariance, raising the questions of \emph{how} and \emph{why} it emerges in biological systems. %Addressing these questions requires both mechanistic and functional perspectives.

Unlike nonliving systems, which are governed primarily by thermodynamics and passive stochastic dynamics, living systems are shaped by evolutionary selection and can optimize a wide range of regulatory objectives \cite{smith1978optimization}. Selection for functional traits can drive biological systems far from the neutral, random, or typical expectation~\cite{hledik2022accumulation,sokolowski2025deriving}. Therefore, understanding the origins and implications of the observed $\tc$-invariance requires examining both the molecular mechanisms that regulate transcription (the ``how'') and the potential functional benefits it confers (the ``why'').

Recent studies have begun to integrate mechanistic models inferred from data with theoretical frameworks that describe regulatory architectures in terms of optimization principles, selectable quantitative phenotypes, and physical constraints~\cite{Wiktor:2021,Zoller:2022,tkavcik2025information}. In this context, we view transcriptional regulation as balancing trade-offs between molecular resource availability, reaction speeds, and energy dissipation~\cite{Tkacik:2008,Govern:2014,Grah:2020} to achieve rapid yet accurate gene expression control. Such trade-offs may necessitate energy dissipation and force transcriptional regulation into a non-equilibrium regime. 

Theoretical models suggest that non-equilibrium dynamics can enhance regulatory precision and robustness, and mitigate conflicting signals, such as those arising from crosstalk ~\cite{cepeda2015stochastic,Estrada:2016ct, Scholes:2016gz, Grah:2020,Zoller:2022,Lammers:2023,Martinez:2024,perkins2025chromatin}. Yet, whether eukaryotic transcription actively exploits these advantages remains an open empirical question~\cite{Suter:2011hk, Zoller:2015im, Li:2018ha, wong2020gene,Shelansky:2022}. Similarly, it is unclear whether $\tc$-invariance necessarily implies non-equilibrium function, or whether the two properties are causally linked at all.

Here, we identify the minimal mechanistic model required to reproduce $\tc$-invariance and reveal its functional significance. We show that a constant $\tc$ naturally emerges from a regulatory cycle that must necessarily be out of equilibrium. In this framework, a balance between transcriptional facilitation and stabilization by an activated enhancer maintains invariant switching dynamics at the promoter. We show that this invariance is not a generic outcome of biochemical models but instead likely reflects an evolutionary signature of selection for maximal information flow given limited promoter switching speed. 

\section*{Results}

\subsection*{Empirical constraints for models of transcriptional regulation}

%TC:ignore
\begin{figure*}
\centering
\includegraphics[scale=1.25]{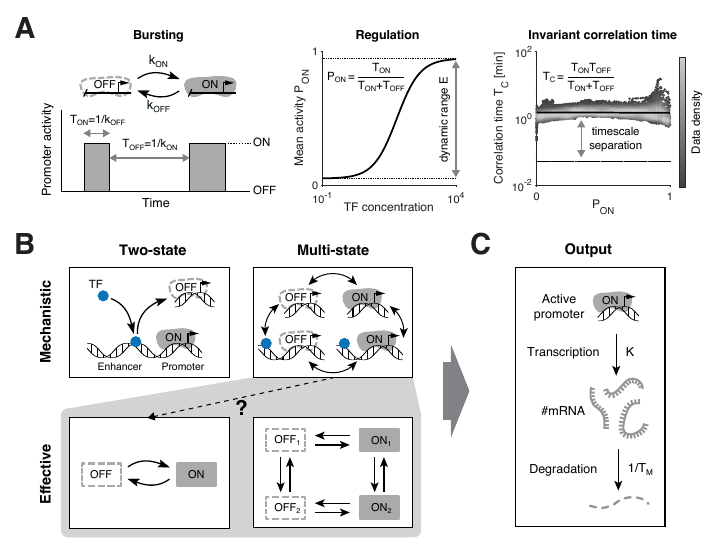}
\caption{{\bf Empirical constraints inform the structure of minimal transcriptional regulatory models.}
{\bf (A)} Experimental constraints include: \emph{(i)} bursting, i.e., stochastic switching between transcriptionally inactive (OFF) and active (ON) promoter states, with mean durations $T_{\rm OFF}$ and $T_{\rm ON}$ (left); \emph{(ii)} regulation, quantified by the induction curve relating the mean activity $P_{\rm ON}=T_{\rm ON}/(T_{\rm ON}+T_{\rm OFF})$ to TF concentration, spanning a large dynamic range $E$ (middle); and \emph{(iii)} a nearly constant promoter switching correlation time $\tc = T_{\rm OFF} P_{\rm ON}$, invariant across expression levels and significantly slower than the typical (dotted line) TF residence time (right).
{\bf (B)} Conceptual classification of regulatory models: two-state vs. multi-state (horizontal), and mechanistic vs. effective (vertical). Mechanistic two-state models (top left) cannot reproduce invariant $\tc$ (panel A, right). Effective two-state models (bottom left) accurately describe data; our goal is to find mechanistic multi-state models (top right) that coarse-grain to such effective models and reproduce $\tc$-invariace (dashed arrow).
{\bf (C)} ON/OFF promoter switching can be coupled to transcriptional output via a birth-death process, generating experimentally measurable mRNA distributions in single cells.}
\label{fig1}
\end{figure*}
%TC:endignore

Motivated by the surprising invariance of $\tc$ and its potential functional significance, we aim to identify the simplest regulatory architectures that can account for this phenomenon. Concretely, we focus on mechanistic models of transcriptional regulation consistent with three core experimental findings. Because they place stringent constraints on the underlying models, we will refer to these experimental findings, which we enumerate below and summarize in Fig.~\ref{fig1}A, as ``empirical  constraints'' throughout the remainder of this work. 

First, transcription occurs in bursts: promoters stochastically alternate between active (ON) and inactive (OFF) states, with average durations $T_{\rm ON}$ and $T_{\rm OFF}$, respectively. Second, the average promoter activity, $P_{\rm ON} = T_{\rm ON}/(T_{\rm ON}+T_{\rm OFF})$ is modulated by TF concentrations across a large dynamic range, $E=(\max P_{\rm ON})/(\min P_{\rm ON})$. Third, the switching correlation time, $\tc = T_{\rm OFF} P_{\rm ON} \sim 1.5$ min remains approximately constant over the full range of $P_{\rm ON}$~\cite{Chen:2023}, and is typically much longer than the TFs residence time on the DNA~\cite{Jonge:2022}. 

This third constraint is particularly striking. Unlike a bound on a parameter, it specifies an entire functional relationship between $\tc$ and $P_{\rm ON}$ that robustly holds across conditions and many genetic perturbations. Published models currently neither explain the mechanistic origin of this invariance, nor do they clarify the conditions under which it can arise.

To evaluate candidate models, we represent transcriptional regulation as a continuous-time Markov process, where promoter activity $P_{\rm ON}$ corresponds to the stationary probability of being in an active, transcribing configuration. As a first step, we consider minimal models built from elementary reactions such as unimolecular decays or bimolecular interactions (e.g., TF-DNA binding/unbinding; see Fig.~\ref{figS1}). These models can reproduce transcriptional bursts and tunable $P_{\rm ON}$, but not an invariant $\tc$~(see SI~Appendix Sec 3.1). In these models only one rate (e.g., $k_{\rm ON}$ or $k_{\rm OFF}$) depends on TF concentration. Achieving constant $\tc$ would require finely tuned, anti-correlated adjustments to both $T_{\rm ON}=1/k_{\rm OFF}$ and $T_{\rm OFF}=1/k_{\rm ON}$ simultaneously, a scenario which is biologically implausible.

We then consider mechanistic multi-state models, particularly those based on cooperative TF binding~\cite{Estrada:2016ct,Martinez:2024}, which introduce additional intermediate states (Fig.~\ref{figS2}; see SI Appendix Sec. 3.2). These models can approximate constant $\tc$ over part of the expression range, but they fall short in two important ways. First, they predict $\tc = 1/b$, where $b$ is the TF unbinding rate -- incorrectly tying promoter switching to TF residence time on the DNA. This contradicts observations that switching occurs on a much slower, TF-independent timescale~\cite{Estrada:2016ct,Martinez:2024}. Second, they give rise to multiple promoter switching correlation timescales, while data support a single dominant $\tc$. Thus, while $\tc$ can be formally calculated for this model class, it lacks a clear physical meaning, since it arises from a mix of latent dynamics that cannot be coarse-grained into an effective two-state model.

To overcome these issues, we adopt a systematic, constraints-based approach. Rather than carefully hand-crafting a single model build out of plausible molecular interactions, we explore a broad space of multi-state reaction networks to identify those consistent with empirical constraints~\cite{Zoller:2022}. We organize these models along two independent axes~(Fig.~\ref{fig1}B). Horizontally, we distinguish two-state from multi-state models. Vertically, we separate mechanistic models -- built from explicit chemical microstates -- from effective models, which provide coarse-grained descriptions and abstract away molecular detail. Effective models can often be identified and tractably inferred from data but lack physical interpretability.

Previous studies show that two-state effective models provide excellent minimal descriptions of transcriptional bursting~\cite{Zoller:2018gj,Chen:2023}. Thus, our goal is to identify minimal mechanistic multi-state models that can be coarse-grained into effective two-state behavior while  satisfying the $\tc$-invariance and other empirical constraints. 

To connect mechanistic and effective models, we define a quantitative criterion for their dynamical similarity. Specifically, we require the noise-filtering function $\phi_N(\tau)$ of a multi-state model to approximate that of an equivalent two-state model $\phi_2(\tau)$ across all timescales $\tau$~\cite{Warren:2006ky,Lestas:2008hy,Tkacik:2008ht}:
\begin{equation}
\Delta = \max_{\tau} \left|\phi_2(\tau)-\phi_N(\tau)\right| \leq \delta.
\label{equ:crit}
\end{equation}
where $\delta\in[0,1]$ sets the maximum tolerable mismatch. Analytical work (SI Appendix Sec 2.2) provides the basis for introducing $\Delta$ as a relevant measure and details various interesting limiting cases. $\delta\ll 1$ ensures that the $T_{\rm ON}$ and $T_{\rm OFF}$ of the multi-state model match those of the two-state model, and that $\tc$ remains the dominant timescale of promoter dynamics, preserving both its interpretability and its experimental relevance. 

Finally, to relate regulatory mechanisms to measurable outputs, we couple the ON/OFF promoter dynamics to a simple birth-death process for mRNA production and degradation~(Fig.~\ref{fig1}C). This allows us to predict the full mRNA distribution in single cells for any TF input. More broadly, this framework enables connections between physical models and functional phenotypes, such as information flow~\cite{tkavcik2008information,Tkacik:2008,rieckh2014noise} and specificity~\cite{Grah:2020,Zoller:2022}; as well as bridging theory and experiment in a way that exposes the constraints and trade-offs of transcriptional regulation.

\subsection*{A minimal non-equilibrium cycle model captures $\tc$-invariance}

%TC:ignore
\begin{figure*}
\centering
\includegraphics[scale=1.25]{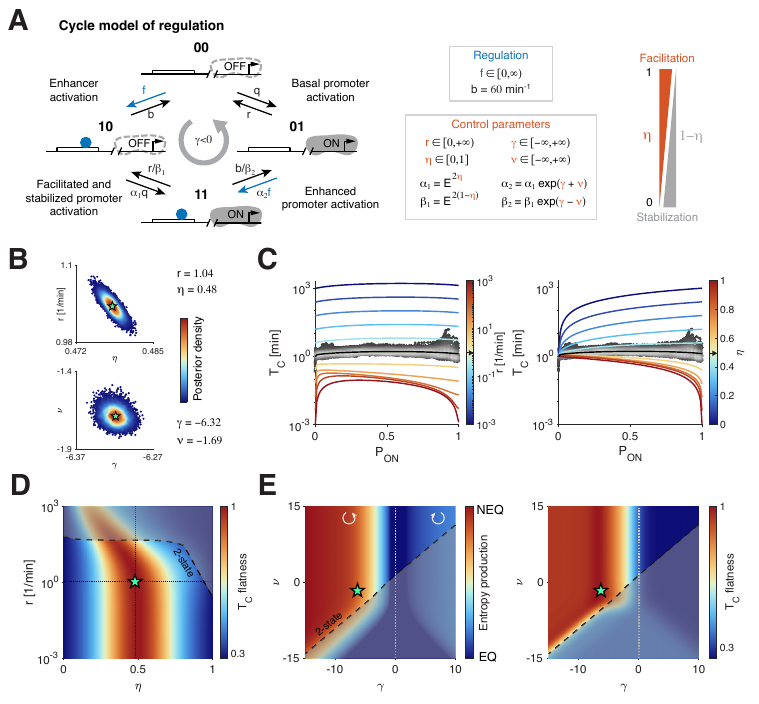}
\caption{{\bf A minimal non-equilibrium model captures $\tc$-invariance  and reveals functional constraints.}
{\bf (A)} Schematic of the four-state ``cycle model'' of transcriptional regulation. A possible mechanistic interpretation of the model shown here assumes that TF binding (blue, rate $f$) modulates enhancer activation to affect promoter state. States are labeled by enhancer (first binary digit) and promoter (second binary digit) activation. Transcription occurs in states with promoter ON (gray).  Reaction rates combine compactly into the following control parameters: $b$ (TF unbinding rate), $r$ (basal promoter deactivation rate), $\eta$ (regulatory tradeoff between stabilization and facilitation), $\gamma$ (cycle directionality), and $\nu$ (activation branch asymmetry). 
{\bf (B)} Posterior distributions of model parameters obtained by fitting the cycle model  $\tc(P_{\rm ON})$ measurements in \textit{Drosophila} embryos~\cite{Chen:2023}. Two 2D projections of the 4D posterior are shown, with a cyan star marking the best-fit values (posterior mean). We used Gaussian likelihood in log-space,  uninformative priors, and empirical errors from Fig.~\ref{fig1}.
{\bf (C)} Systematic parameter variations (colorbar) around the best-fit value: $\tc$ shape is insensitive to changes in $r$ (left), but strongly deformed by small changes in $\eta$ (right). Experimental $\tc$ data from Fig.~\ref{fig1}A shown in gray for reference.
{\bf (D)} Landscape of $\tc$ flatness across $(r,\eta)$ plane (see also SI Appendix Sec 5.6). $\tc$-invariance (i.e., flatness $\approx 1$) emerges only within a narrow regime of $\eta\approx 0.5$ and $r \lesssim b$, where enhancer facilitation and stabilization are balanced and time-scale separation is observed. Cyan star: best-fit values from panel B. Semi-transparent region: models that (unlike data) cannot be coarse-grained to a two-state description according to Eq.~(\ref{equ:crit}).
{\bf (E)} Effect of $(\gamma, \nu)$ on dissipation (left) and $\tc$ flatness (right). $\gamma = 0$ (vertical dotted line) defines the thermodynamic equilibrium manifold; models with $\gamma \neq 0$ are non-equilibrium. White circular arrows: probability flow direction in the cycle. High $\tc$ flatness is confined to the $\gamma < 0$ region within the coarse-grainable regime.  Models with $\gamma \gg 0$ produce less bursty dynamics and fail to match  experimental constraints.
}
\label{fig2}
\end{figure*}
%TC:endignore

To gain intuition, we started by looking for the simplest biophysically plausible mechanistic model composed of elementary reactions that maintains a constant switching correlation time $\tc$ (Fig.~\ref{figS3}A). We constructed this model incrementally. First, we identified two separate two-state models, each exhibiting identical $\tc$ values, but differing in their mean promoter activity (high and low). Next, we coupled the two models by introducing a regulatory element whose binding by a TF drives transitions between the two (high and low) promoter activity regimes. These transitions are rapid and concentration-dependent, capturing TF association and dissociation. The resulting four-state mechanistic model allows tunable activity levels (Fig.~\ref{figS3}B), maintains an invariant switching correlation time (Fig.~\ref{figS3}C), and remains effectively two-state from a coarse-grained perspective (Fig.~\ref{figS3}D), according to the criterion of Eq.~(\ref{equ:crit}). It naturally separates two timescales: fast  TF (un)binding and slow promoter switching dynamics (Fig.~\ref{figS3}E), a key feature observed in data.

To explore a broader range of behaviors compatible with $\tc$-invariance, we extended the resulting reaction scheme into a more versatile four-state ``cycle model.'' This model can parametrically accommodate a large space of behaviors, enabling both functional tuning (e.g., of the $\tc(P_{\rm ON})$ curve) and thermodynamic control via dissipation. Similar cycle-based architectures have  been proposed previously in the context of regulatory precision and non-equilibrium control \cite{Mirny:2010,Tkacik:2011jr,Grah:2020,Lammers:2023}. 

A possible mechanistic interpretation of the cycle model is shown in Fig.~\ref{fig2}A: promoter activation and deactivation occur either basally (states {\tt 00} and {\tt 01}, with rates $q$ and $r$, respectively) or under the influence of an active enhancer (states {\tt 10} and {\tt 11}). Enhancer activation is driven by TF binding (transitions between states {\tt 00} and {\tt 10}, with a concentration-dependent rate $f$ and unbinding rate $b$).  Once the enhancer is active, it can both facilitate promoter activation (via $\alpha_1>1$) and stabilize transcriptional activity (via $\beta_1>1$), thereby promoting entry into, as well as residence in, the doubly active state {\tt 11}. 

The cycle model generates typical sigmoidal induction curves (Fig.~\ref{figS4}A), where the probability of being transcriptionally ON, $P_{\rm ON} = P(\mathtt{01})+P(\mathtt{11})$, increases monotonically with the TF binding rate, $f$. The ON probability ranges from $P_{\rm min}=q/(q+r)$ at $f=0$ to $P_{\rm max} = \alpha_1\beta_1 q/(\alpha_1\beta_1 q + r)$ as $f\rightarrow \infty$. To ensure a wide dynamic range of expression consistent with experimental observations~\cite{Chen:2023}, we define the range as $E=P_{\rm max}/P_{\rm min}$ and choose $E=10^{3}$. This constraint implies specific relationships among parameters:  $r=Eq$ and $\alpha_1 \beta_1 = E^2$. 

The model's reaction rates can be reparametrized to make the model more amenable to analysis and interpretation. Below, we introduce the new parameter set $r$, $\eta$, $\gamma$, and $\nu$, which provides direct control over promoter responsiveness and non-equilibrium behavior (see also Fig.~\ref{fig2}A and caption for definitions).

To relate enhancer facilitation and stabilization  to the dynamic range $E$, we introduce a parameter $\eta\in[0,1]$, with $\alpha_1=E^{2\eta}$ and $\beta_1=E^{2(1-\eta)}$. Intuitively, $\eta$ governs the balance between two regulatory mechanisms: when $\eta=0$, the system relies entirely on stabilization of the doubly active state {\tt 11}  ($\alpha=1$ and $\beta_1=E^2$), whereas $\eta=1$ corresponds to pure facilitation of the transition $\mathtt{10}\rightarrow\mathtt{11}$ into the doubly active state  ($\alpha=E^2$ and $\beta_1=1$). Intermediate $\eta$ values tune the relative contributions of each mechanism (see SI Appendix Sec 5.1).

The (non)equilibrium dynamics of the cycle model can be examined by analyzing the stationary probability currents. Detailed balance holds when $\alpha_1\beta_1=\alpha_2\beta_2$, with cycle reversibility controlled by $\alpha_2$ and $\beta_2$ (see SI Appendix Sec 5.3). To systematically explore departures from equilibrium, we reparameterize these rates using two new variables: $\alpha_2=\alpha_1 \exp{(\gamma+\nu)}$ and $\beta_2=\beta_1\exp{(\gamma-\nu)}$. Here, $\gamma \in (-\infty,+\infty)$ controls the system’s dissipation and directionality, while $\nu \in (-\infty, +\infty)$ sets the asymmetry between the two reaction branches. Together, these parameters allow a systematic exploration of the transition between equilibrium ($\gamma=0$) and non-equilibrium ($\gamma\neq0$) regimes.

In sum, the model's behavior depends on four free control parameters: $r,\eta,\gamma,\nu$. Other parameters are fixed by empirical constraints. For example, the TF-dependent rate $f$ must remain tunable to dial the mean promoter activity across the entire dynamic range,  $E=10^3$. TF unbinding rate $b$ is fixed to $b=60\,$min$^{-1}$, in line with experimentally measured TF residence times on the DNA~\cite{Jonge:2022} (a ten-fold lower $b$ leads to higher $\nu$ but leaves other inferred parameters as well as all our conclusions unchanged). $b$ also sets the absolute timescales for the cycle model. 

Fitting the cycle model to the empirical $\tc$ function extracted from \emph{Drosophila} transcriptional data~\cite{Chen:2023} yields well-constrained parameters (Fig.~\ref{fig2}B). The best-fit parameter values indicate clear non-equilibrium behavior, with $\gamma < 0$. Systematic variation of the model's parameters shows that $r$ primarily controls the scale of $\tc$ (its average value), while $\eta$ modulates its shape (Fig.~\ref{fig2}C). Invariance of $\tc$ -- i.e., a constant, flat $\tc$ as a function of $P_{\rm ON}$ -- is only maintained within a narrow parameter range:  $\eta\sim 0.5$ and $r \lesssim b$ (Fig.~\ref{fig2}D). These constraints are also necessary for the observed time-scale separation between TF binding and enhancer switching; they lead to  a precise balance between enhancer-mediated facilitation and stabilization of the promoter. At $\eta=0.5$, the enhancer mirrors the reverse kinetics of basal promoter activation, with $\alpha_1 = \beta_1 = E$, so that $\alpha_1 q = r$ and $r/\beta_1 = q$. While the best-fit model remains accurately coarse-grainable into an effective two-state description, moderate deviations in parameters can break this property, leading to qualitatively different behavior where $\tc$ loses its interpretability (semi-transparent region in Fig.~\ref{fig2}D, see also Fig.~\ref{figS4}G).

To understand when coarse-graining is feasible, we analytically identified the conditions under which the cycle model satisfies the the two-state criterion of Eq.~(\ref{equ:crit}) exactly, i.e., with $\delta=0$ (see SI Appendix Sec 5.8). Surprisingly, we demonstrate that perfect coarse-graining into an effective two-state model is possible precisely when $\tc$ is strictly invariant and equal to $1/(r+q)$, consistent with the plateau observed in Fig.~\ref{fig2}C. This analysis clarifies the mechanistic requirements for $\tc$-invariance and links them to the elementary rates of the underlying reaction network.

The cycle model can be extended to include cooperative TF binding (Fig.~\ref{figS5}; SI Appendix Sec 5.8), which, however,  does not change our qualitative conclusions. Attempts to reduce the four-state cycle model to a simpler, three-state system fail to reproduce the desired $\tc$-invariance (Fig.~\ref{figS6}). Consequently, the four-state cycle model is  the simplest mechanistically plausible reaction scheme consistent with empirical constraints.

Does the observed $\tc$-invariance require  out-of-equilibrium dynamics? To address this question, we varied $\gamma$ and $\nu$ over a broad range around their best-fit values, thereby scanning through both the equilibrium ($\gamma=0$, or $\nu \rightarrow -\infty$) and the non-equilibrium regimes. As $\gamma$ is displaced from zero, dissipation increases (Fig.~\ref{fig2}E, left), and the net probability flux reverses direction upon crossing the equilibrium boundary (SI Appendix Sec 5.3-5.4). Importantly, $\tc$-invariance cannot be achieved anywhere along the equilibrium parameter manifold (Fig.~\ref{fig2}E, right), a result we prove analytically for all $r$ and $\eta$ within the four-state cycle model class (see SI Appendix Sec 5.7). The converse does not hold: not all non-equilibrium models exhibit $\tc$-invariance. Taken together, our results rule out thermodynamic equilibrium as a viable operating regime for models exhibiting invariant $\tc$. This motivates us to ask whether there are any  functional benefits that justify the energetic cost for maintaining  the system out of equilibrium.

\subsection*{Invariant $\tc$ is a signature of optimized information flow}

%TC:ignore
\begin{figure*}
\centering
\includegraphics[scale=1.25]{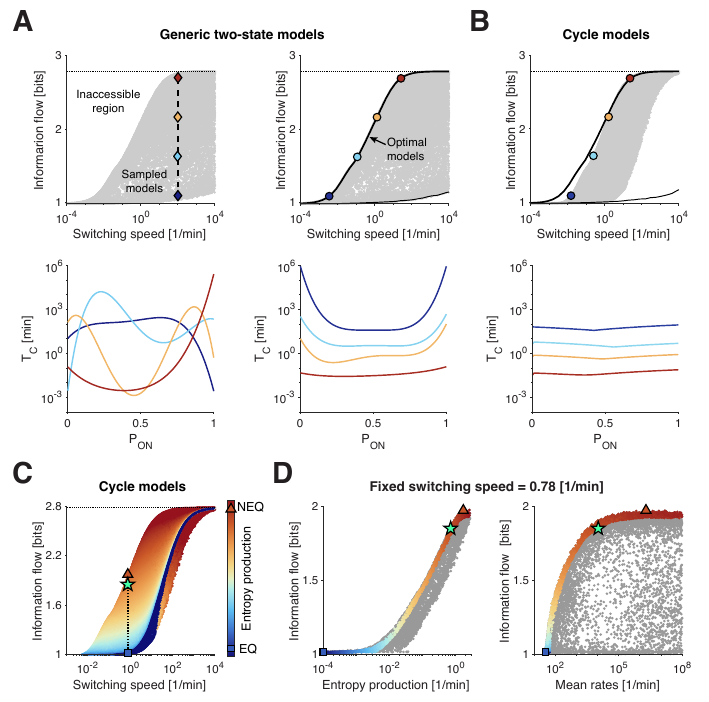}
\caption{{\bf Optimizing information flow under a switching speed constraint predicts invariant $\tc$.}
{\bf (A)} Top: Information flow $I$ as a function of switching speed $V$, sampled for effective two-state models with diverse $\tc(P_{\rm ON})$ functions (gray dots). Maximal information flow (dotted line) is achieved in the Poisson limit ($V\rightarrow\infty$). Four representative models at fixed $V$ are highlighted (colored diamonds; top left). Models that maximize $I$ at fixed $V$ (colored circles; top right) lie on the Pareto front (upper black line). Bottom: Corresponding $\tc$ functions for the highlighted models.
{\bf (B)} Top: Manifold of coarse-grainable four-state cycle models (gray dots) plotted in the $I$–$V$ plane. Cycle models, particularly at high $I$, closely approach the effective two-state Pareto front (black line, reproduced from A). Bottom: Cycle models on the Pareto front exhibit nearly invariant $\tc$, in contrast to off-the-front models.
{\bf (C)} Information flow $I$ vs switching speed $V$ for cycle models, colored by minimal dissipation (entropy production) required to achieve the corresponding $(V,I)$ combination. Equilibrium models (dark blue) lie far below the Pareto front, whereas non-equilibrium models (brighter colors) achieve higher $I$ at lower $V$. The best-fit model from Fig.~2B (cyan star, $V^* \sim0.78 \text{min}^{-1}$) lies near the front. Matched $V^*$ equilibrium and non-equilibrium models are highlighted (blue square and red triangle, respectively). 
{\bf (D)} Sections of the regulatory phenotype manifold for cycle models at fixed switching speed $V^* \sim0.78 \text{min}^{-1}$ (gray dots, as in B). The Pareto front (colored dots, as in C) tracks trade-offs between information flow, energy dissipation, and mean reaction rates.  The same three models from C are highlighted.
}
\label{fig3}
\end{figure*}
%TC:endignore

Gene activity controls mRNA levels, which represent a temporal average over stochastic promoter ON/OFF transitions. This average is taken over a fixed timescale, the mRNA lifetime. If $\tc$ remains invariant across regulatory conditions, mRNA levels will always integrate and hence average over the same effective number of promoter transitions, each encoding information about upstream TF concentrations. This suggests a qualitative view of an efficient information transmission pipeline, in which the reaction network is tuned to maintain a consistently high precision across the full dynamic range of inputs. But does this intuition hold up quantitatively?

To examine this idea quantitatively, we computed the information flow through different regulatory models. The input to each model was the control parameter $P_{\rm ON}$, which reflects promoter activation probability and can equivalently be mapped to the kinetic parameter $f$ or the TF concentration. The output was the mRNA copy number $M$, a random variable with a steady-state distribution given by the corresponding birth-death process (Fig.~\ref{fig1}C and Fig.~\ref{figS7}A, assuming a constant transcription initiation rate $K$ and a fixed mRNA lifetime $T_{\rm M}$). Together, these define an information channel, $p(M|P_{\rm ON})$, which can be fully characterized by solving the corresponding Master Equation (see SI Appendix Sec 6.1-2). ``Information flow'' can then be defined for each model as the channel capacity, i.e., the maximum mutual information $I(M;P_{\rm ON})$, achievable by optimizing the input distribution $p^*(P_{\rm ON})$~\cite{Blahut:1972}. The resulting information flow, $I$, is our functional metric for each regulatory model.

Within the class of effective two-state models, the shape of the $\tc(P_{\rm ON})$ function fully determines the model's behavior. We can therefore ask which functional forms of $\tc$ optimize information flow. Unsurprisingly, we find that information is maximized as  $\langle \tc \rangle \rightarrow 0$, where promoter switching becomes infinitely fast. In this so-called ``Poisson limit,'' burstiness (and therefore promoter switching noise) vanishes~\cite{tkavcik2008information}, and the promoter effectively behaves as if it were continuously transcribing at rate $KP_{\rm ON}$. The only remaining source of noise in the channel is then the stochastic birth-death dynamics of mRNA.

Physical reaction rates in any real system are necessarily finite, ruling out the Poisson limit and imposing a constraint on promoter switching speed. To capture this constraint quantitatively, we define the effective switching speed as $V=\avg{1/\tc}=\int dP_{\rm ON}\;p^*(P_{\rm ON})/\tc(P_{\rm ON})$. As expected, $V$ diverges in the Poisson limit, highlighting its idealized and unphysical nature. 

For all other effective two-state models, we broadly sample the space of admissible $\tc(P_{\rm ON})$ functions (see SI Appendix Sec 6.3), and for each sampled model, we compute two key phenotypes: the information flow $I$ and the switching speed $V$ (Fig.~\ref{figS7}B). Plotting these quantities against one another reveals a structured manifold bounded from above and below (Fig.~\ref{fig3}A). The lower boundary corresponds to inefficient models with minimal information flow given a fixed switching speed -- this boundary is not biologically relevant. In contrast, the upper boundary defines a \emph{bona fide} Pareto front: information flow can only be increased  by allowing faster switching. As $V\rightarrow\infty$, $I$ approaches the Poisson limit. Strikingly, models on this optimal front exhibit increasingly invariant $\tc$ functions as information flow increases, providing the first quantitative link between information flow optimality and $\tc$-invariance. While broadly consistent with the intuition that invariant $\tc$ ensures uniform temporal averaging over promoter noise across the full induction range, the mathematical emergence of this connection is both non-trivial and highly unexpected.

What region of the optimal manifold is accessible to our mechanistic cycle models, and can they approach the Pareto front? To address this, we broadly sampled the four independent model parameters ($r,\eta,\gamma,\nu$), retaining only those configurations that can be reliably coarse-grained into an effective two-state description ($\delta<0.1$ in Eq.~(\ref{equ:crit})). The resulting cycle models densely populate the phenotype space and tightly approach the Pareto front defined by effective two-state models (Fig.~\ref{fig3}B): the tighter, the higher the allowed switching speed. Importantly, cycle models that maximize information flow along the Pareto front consistently exhibit invariant $\tc$ across the full range of $V$; in contrast, $\tc$-invariance disappears away from the front. These findings suggest that $\tc$-invariance is not a generic feature, but rather a specific consequence of optimizing information flow under  kinetic constraints.

We next examined how the information flow $I$ and switching speed $V$ in cycle models relate to entropy production (or, equivalently, energy dissipation) $\dot{S}$. As shown in Fig.~\ref{fig3}C, the entire equilibrium manifold, comprising models with $\gamma=0$, falls substantially below the Pareto front and, by extension, well below the best performing non-equilibrium models~(see also Fig.~\ref{figS8}A). Compared to their equilibrium counterparts, non-equilibrium models can support significantly higher information transmission while operating at slower switching speeds. These differences are far from negligible: in some cases, gains in information can reach nearly 1 bit, highlighting a substantial functional benefit of non-equilibrium regulation. Importantly, the best-fit non-equilibrium cycle model from Fig.~\ref{fig2}B lies very close to the Pareto front, providing clear support for the information optimization hypothesis in the context of \emph{Drosophila} gap gene regulation.

The analysis of constraints and trade-offs can be extended further (Fig.~\ref{figS9}A). We expect that the true Pareto front for cycle models lives in a high-dimensional space of regulatory phenotypes. While Fig.~\ref{fig3}A explored the projection of these phenotypes onto the $I$ vs $V$ plane, other sections of the solution space may yield complementary insights. 

To illustrate this, Fig.~\ref{fig3}D fixes the switching speed to the value inferred from data ($V^* \sim0.78 \text{min}^{-1}$) and displays two Pareto front cross-sections. Along the front, the cycle model can be continuously ``deformed'' from best equilibrium model (no dissipation, $I=1$ bit) to the best non-equilibrium model (maximal dissipation, $I\approx 2$  bits). This increase in performance, however, comes at a steep cost: achieving maximal information flow requires the mean reaction rates in the cycle to increase by nearly four orders of magnitude. 

Interestingly, while the best-fit system lies close to the optimal non-equilibrium model in terms of information flow and energy dissipation (requiring a $\sim 2.4$ fold increase in entropy production to close an information gap of $\approx 0.12$ bits), it achieves this performance with mean rates that are two orders of magnitude lower. This contrast suggests that biological systems may face important constraints not only on energy usage but also on the magnitude of individual reaction rates, a consideration that has received surprisingly little attention to date.

In line with the perspective laid out in the introduction, our analysis integrates prior empirical observations, data-driven inference, physico-chemical constraints, and optimization theory to delineate the space of regulatory models that are both admissible and functional. Within this space, a simple class of four-state cycle models emerges as particularly instructive. These models not only reproduce key features of the data but also reveal non-trivial trade-offs that biological systems must navigate. Our results suggest that selection for regulatory precision may have driven the emergence of invariant $\tc$. This invariance, near the Pareto front of functional performance, is as a hallmark of optimal information transmission given physical constraints.

\subsection*{Cycle models are robust to perturbations}

%TC:ignore
\begin{figure*}
\centering
\includegraphics[scale=1.25]{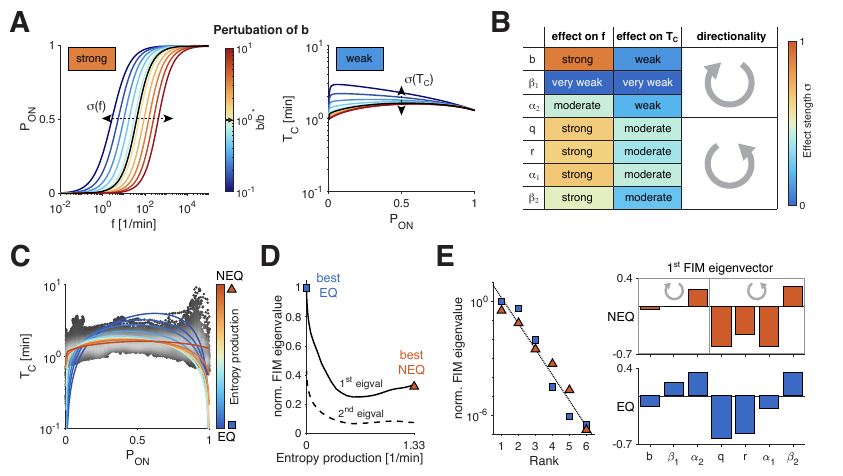}
\caption{{\bf Best-fit cycle model is robust to perturbations.}
{\bf (A)} Perturbing the TF unbinding rate $b$ by $\pm 10$-fold (log-normal noise around the best-fit value $b^*$) has a strong effect on the induction curves $P_{\rm ON}(f)$ (left), but only a weak effect on the switching correlation time $\tc(P_{\rm ON})$, consistent with experimental findings.
{\bf (B)} Systematic analysis of individual parameter perturbations reveals that rates aligned with the natural counter-clockwise direction of the transcription cycle (e.g., $r$, $q$, and $\alpha_1$) have strong effects on induction and moderate effects on $\tc$, whereas reverse-direction rates have minimal impact on $\tc$. Color scale quantifies effect strength (standard deviation across perturbations); gray arrows indicate cycle directionality aligned with the perturbed rates. See also Fig.~\ref{figS10}A--B for quantitative effect sizes.
{\bf (C)} $\tc$ functions for a family of interpolating models, generated by varying the entropy production (color scale, controlled by $\gamma$) between the best-fit NEQ model ($\gamma=-6.32$, dark orange; cyan star in Figs.~\ref{fig2} and \ref{fig3}) and the best-fit equilibrium model (EQ) ($\gamma=0$, dark blue). For each $\gamma$, all remaining parameters were refit to the observed $\tc$. Fit quality steadily deteriorates toward equilibrium. See also Fig.~\ref{figS10}C–E for parameter values and refitting performance.
{\bf (D)} First two eigenvalues of the Fisher Information Matrix (FIM) as a function of entropy production. Eigenvalues are normalized to the largest eigenvalue for the best-fit EQ model. 
{\bf (E)} Normalized eigenvalue spectra (left) of the best-fit NEQ and EQ models reveal ``sloppiness'', a logarithmic hierarchy of sensitivities. The dominant eigenvector (right) for the NEQ model shows that sensitivity is concentrated in parameters aligned with cycle direction ($q$, $r$, and $\alpha_1$), whereas the EQ model displays more diffuse and unfocused sensitivity.}
\label{fig4}
\end{figure*}
%TC:endignore

Perturbation experiments in \emph{Drosophila} gap genes have revealed a striking robustness of $\tc$-invariance~\cite{Chen:2023}. Changes in TF concentrations or mutations/deletions in enhancer sequences primarily affected the mean gene activity ($P_{\rm ON}$) but only had a limited impact on the magnitude or shape of $\tc$. Many such cis- and trans-perturbations, ranging from altered TF levels and chromatin modifications to mutations in DNA binding sites or enhancer deletions, are expected to affect the ``enhancer activation'' step of the regulatory cycle (Fig.~\ref{fig2}A). In our model, these perturbations are effectively captured by changes in the TF unbinding rate ($b$). When $b$ is perturbed tenfold around its optimal value, the induction curves $P_{\rm ON}(f)$ exhibit substantial shifts, yet the $\tc$ remains largely unchanged (Fig.~\ref{fig4}A), consistent with the experimental observations~\cite{Chen:2023}.

To systematically evaluate whether this robustness is recapitulated by the cycle model, we perturbed each of its elementary reaction rate parameters ($b, q, r, \alpha_1, \beta_1, \alpha_2, \beta_2$; see Fig.~\ref{fig2}A), one at-a-time, around their best-fit values, by introducing multiplicative log-normal noise. This approach, equivalent to additive Gaussian noise in log-space, is well-justified, as kinetic rates are strictly positive and naturally defined on a multiplicative scale. We chose to perturb elementary reaction rates rather than the derived control parameters ($r, \eta,\gamma,\nu$), which are more convenient for analytical calculations, because the former have a clearer mechanistic interpretation and can be more directly linked to mutational effects. For each perturbed model, we quantified changes in the induction and the $\tc$ curves, by computing the standard deviation of the logarithmic mean across perturbations (SI Appendix, Sec 7.1). The largest effects on induction were caused by perturbing $b$ (TF unbinding), $r$ and $q$ (promoter (de)activation), and $\alpha_1$ (enhancer facilitation), as shown in  Figs.~\ref{fig4}B and \ref{figS10}A. In contrast, $\tc$ was more robust overall, with only moderate sensitivity to changes in $r$, $q$, $\alpha_1$, and $\beta_2$, as shown in Figs.~\ref{fig4}B and \ref{figS10}B. These model predictions qualitatively mirror the robustness patterns observed experimentally.

Could robustness be a general feature of non-equilibrium dynamics? The parameters that most strongly influence $\tc$ in our perturbation analysis ($q$, $r$, $\alpha_1$, $\beta_2$) correspond to reaction steps aligned with the natural, counter-clockwise progression of the transcriptional cycle (Fig.~\ref{fig4}B). In contrast, reactions oriented opposite to this direction have negligible influence on $\tc$. This directional asymmetry in sensitivity suggests that non-equilibrium dynamics---by favoring a unidirectional flow through the transcriptional cycle---may inherently buffer the system against a broad class of local perturbations.

To investigate this idea more systematically, we assessed how robustness varies across the model’s parameter space as a function of entropy production. Starting with our best-fit non-equilibrium model (NEQ; with $\gamma=-6.32$), we generated a family of interpolating models by gradually increasing $\gamma$ toward zero, thereby reducing entropy production and approaching equilibrium (Figs.~\ref{figS10}C–E). For each $\gamma$, all remaining parameters were refit to best match the invariant $\tc$  (Fig.~\ref{fig4}C). As expected, fit quality deteriorated steadily with decreasing dissipation. Nevertheless, this family of models provided a consistent basis for computing the Fisher Information Matrix (FIM), which quantifies the local sensitivity of the $\tc$ fit to changes in the elementary reaction rates (SI Appendix Sec 7.2). 

First, we observed that the leading eigenvalue of the FIM decreased as the best-fit NEQ model was approached and entropy production increased. Specifically, the dominant eigenvalue was approximately threefold higher for the best-fit EQ model compared to the best-fit  NEQ model. This indicates that the $\tc$ function for the EQ model is significantly more sensitive, thus less robust, to coordinated parameter perturbations, reinforcing the single-parameter perturbation results shown in Fig.~\ref{fig4}B. 

Second, we analyzed the full eigenvalue spectrum of the FIM, for both the best NEQ and EQ models. In both cases, the spectrum displayed a characteristic hierarchy, with eigenvalues approximately equally spaced on logarithmic scale: a hallmark of the so-called ``sloppy'' models (Fig.~\ref{fig4}E, left)~\cite{Transtrum:2015}. This indicated that only a few effective parameter combinations are tightly constrained by the $\tc$ data. For the best NEQ model, the eigenvector associated with the leading eigenvalue revealed that three parameters ($q$, $r$, $\alpha_1$) dominate the sensitivity of the $\tc$ function (Fig.~\ref{fig4}E, right). These parameters align precisely with the natural counter-clockwise progression of the transcriptional cycle. In contrast, the best EQ model displayed more broadly distributed and less focused sensitivity to parameter variation. 

Taken together, these results highlight an additional key feature of the non-equilibrium regime: it enables the transcriptional cycle  to maintain high information flow while being robust to multiple perturbation directions in the parameter space. Interestingly, this robustness is not a consequence of a given regulatory ``cycle'' topology \emph{per se}, but rather a consequence of a specific parameter regime within which the cycle model operates.

\section*{Discussion}

This study provides both a mechanistic and functional account of an empirically robust yet theoretically unexplained observation: the invariance of promoter switching correlation time, $\tc$, across a wide range of gene expression levels in early \emph{Drosophila} development. This invariance has now been observed in multiple genes and appears conserved across species, though with varying degrees of measurement precision. Canonical models of transcriptional regulation, including standard two-state models, cannot naturally account for this property. Using simulation and analytical tools, we demonstrate that reproducing $\tc$-invariance requires regulatory architectures with at least four promoter states and broken detailed balance, i.e., systems operating out of thermodynamic equilibrium. These non-equilibrium dynamics unfold on timescales slower than those of individual TF binding and unbinding events~\cite{Grah:2020}, in agreement with experimental observations. Given the growing interest in the equilibrium vs. non-equilibrium nature of gene regulation~\cite{wong2020gene,Zoller:2022}, our identification of $\tc$ invariance as a defining experimental signature provides a concrete, quantitative foothold for further theoretical and empirical exploration.

Four-state transcriptional models are not new: they have appeared in early work on gene expression noise in yeast~\cite{blake2003noise} and are now being revisited using time-resolved single-cell methods~\cite{shelansky2024single}. Theoretical studies have explored the rich phenotypic behaviors that such models enable, including non-equilibrium control of gene regulation~\cite{Grah:2020,Zoller:2022,mahdavi2024flexibility}. Related models have also been linked to mammalian transcriptional data, albeit from different conceptual perspectives~\cite{tunnermann2025enhancer}. This convergence—across systems, methods, and motivations—suggests that a unifying quantitative framework for eukaryotic transcriptional regulation may be emerging, potentially playing a role analogous to that of thermodynamic models in prokaryotic systems.

Cycle models tuned to reproduce $\tc$-invariance exhibit multiple desirable properties. {\bf First}, they can be accurately coarse-grained into effective two-state models that reproduce experimental data. This is of practical importance: two-state models are easier to infer from data, while inference of full mechanistic multi-state models remains challenging and often ill-posed. Yet, when two-state models fit well, they can be interpreted as statistical summaries that compress more complex underlying mechanisms. In this setting, theoretical analysis, not inference, can bridge effective and mechanistic descriptions. We provide an example of such mapping by showing how $\tc$-invariance emerges from coarse-grainable models with broken detailed balance, enabling concrete inferences about non-equilibrium regulatory mechanisms. This mechanistic-to-effective connection also opens new directions for analyzing other biological signal-processing systems with defined input and output states, extending beyond transcription~\cite{reinhardt2023path,tkavcik2025information}.

{\bf Second}, non-equilibrium cycle models exhibit increased robustness to parameter perturbations relative to equilibrium models. Cycle models are particularly robust to changes in parameters controlling reactions that run against the probability flux in the transcriptional cycle. While robustness might be desirable on general grounds, it is especially relevant here, since previously reported genetic perturbations in \emph{Drosophila} failed to disrupt the $\tc$-invariance despite affecting expression levels and induction curves~\cite{Chen:2023}. We also report on the broad eigenvalue spectrum, or sloppiness, of the Fisher Information Matrix~\cite{Transtrum:2015,machta2013parameter}. In our case, sloppiness arises in a model class where both data fitting and  optimization for a functional phenotype (information transmission) are tightly interlinked. Whether the observed sloppiness results primarily from data-fitting-related or from optimality-related degeneracy remains unclear. Moreover, the relevance of sloppiness may depend on whether one analyzes elementary reaction rates or coarse-grained combinations. Systems, such as ours, where optimization, sloppiness, and inference can be analyzed within a single framework, offer an exciting platform for further  explorations~\cite{Wiktor:2021,bauer2025optimization,sokolowski2025deriving}.

{\bf Third}, we find that $\tc$-invariance in constrained cycle models is a direct signature of optimal information transmission. Biologically, this is a statement about intrinsic noise control: maximal information transmission implies the most reliable propagation of signals from transcription factor concentrations to downstream gene expression, given fundamental sources of noise such as promoter switching and stochastic mRNA transcription~\cite{tkavcik2008information,rieckh2014noise}. This connects to classical problems in chemoreception~\cite{Govern:2014,malaguti2021theory}, with the key distinction here being the constraint of $\tc$-invariance. We find that four-state cycle models maximizing steady-state information flow inevitably exhibit this invariance. While we support this numerically and provide intuitive arguments, a general analytical proof remains elusive. 

Importantly, $\tc$-invariance only emerges when the overall switching speed $V$ is constrained. While the value of $V$ sets the timescale of $\tc$, it does not need to be precisely tuned to generate invariance. Without such a constraint, optimization favors infinitely fast switching and Poissonian mRNA statistics, biologically implausible scenarios. Although physical and energetic limits cap $V$, the appropriate mathematical form of this constraint remains unclear: should the bound be placed on the average switching rate, the maximum reaction rate, or some other network-level property? Moreover, switching speed constraints are distinct from energy dissipation limits. For instance, equilibrium models with fast kinetics can suppress bursting noise and approach the Poisson limit (Fig.~\ref{fig3}C), yet still preserve detailed balance. This suggests that switching speed, independent of entropy production, is an underappreciated constraint that crucially shapes the landscape of feasible regulatory dynamics.

Beyond these highlights, our study reveals a complex trade-off landscape among information flow, switching speed, energy dissipation, and internal reaction constraints. These objectives often cannot be simultaneously optimized, either due to fundamental physical limits or evolutionary constraints. Rather than assume the existence of a single optimal model, we advocate for exploring ensembles of near-optimal solutions. In such ensembles, no model achieves the best value for every objective, but many can simultaneously approach optimal values for several. These ensembles are computationally accessible and can be statistically characterized thanks to recent theoretical advances~\cite{Wiktor:2021,Zoller:2022,bauer2025optimization}.

Computationally, such ensembles can be generated by unbiased sampling (as in Fig.\ref{fig3}A,B) or biased toward functional phenotypes. Biologically, this bias mimics selective preference for functional states that are reachable through evolutionary adaptation\cite{sokolowski2025deriving}. Deep links between statistical physics, information theory, and population genetics connect these ensembles to the strength of natural selection~\cite{sella2005application,hledik2022accumulation,barton2025evolution}, supporting the view that optimality in biology is better understood as a landscape of trade-offs rather than a single global maximum.

As an illustration, our analysis (Fig.~\ref{fig3}D) shows that \emph{near-maximal} information transmission can be achieved with reaction rates two orders of magnitude slower than those required for \emph{maximal} information flow. Given the limits to selection in finite populations, such small phenotypic gains in information may be neither resolvable nor selectively advantageous, especially if they come at the cost of reduced robustness or higher energetic demands. These trade-offs suggest that evolution might favor “good enough” solutions over theoretically optimal ones. Quantitatively linking physics-style constrained optimization to evolutionary fitness models thus offers an exciting direction for future research.

In sum, our findings suggest that eukaryotic transcription leverages non-equilibrium regulation to enhance robustness against perturbations and noise. Energy dissipation plays a central role in enabling high information transmission at biologically realistic switching speeds. In the complex regulatory trade-off landscape, moderate dissipation allows systems to approach a performance plateau—beyond which further gains in precision would require disproportionate increases in energy consumption and reaction speed~\cite{Zoller:2022}. The widespread relevance of transcription, the availability of precise measurements, the presence of fundamental biophysical constraints, and the existence of information-theoretic proxies for function together make gene regulation a particularly fertile ground for integrative studies. As such, it offers a uniquely tractable and meaningful setting for connecting biological function, physical law, and evolutionary dynamics.

\section*{Acknowledgments}
This work was supported by the French National Research Agency (ANR-20-CE12-0028 `ChroDynE' and ANR-23-CE13-0021 `GastruCyp' and ANR-10 LABX-73 `Revive'; all TG), and by funding from the European Research Council (ERC-2023-SyG, `Dynatrans', 101118866, TG and GT). This work was also supported in part by the U.S. National Science Foundation, through the Center for the Physics of Biological Function (PHY-1734030, TG), and by National Institutes of Health Grants R01GM097275, U01DA047730, and U01DK127429 (TG).

%%%%%%%%%%%%%%%%%%%%%% BIBLIOGRAPHY %%%%%%%%%%%%%%%%%%%%%%%

%TC:ignore
\bibliography{main_v1}% Produces the bibliography via BibTeX from .bib-file in same folder.

%TC:endignore
%%%%%%%%%%%%%%%%%%%%%% SUPP FIGURES %%%%%%%%%%%%%%%%%%%%%%%

%TC:ignore
\clearpage
\onecolumngrid
\section*{Supplemental Figures}

\makeatletter 
\renewcommand{\thefigure}{S\@arabic\c@figure}
\renewcommand{\thetable}{S\@arabic\c@table}
\setcounter{figure}{0}

\begin{figure*}[h!]
\centering
\includegraphics[scale=1.25]{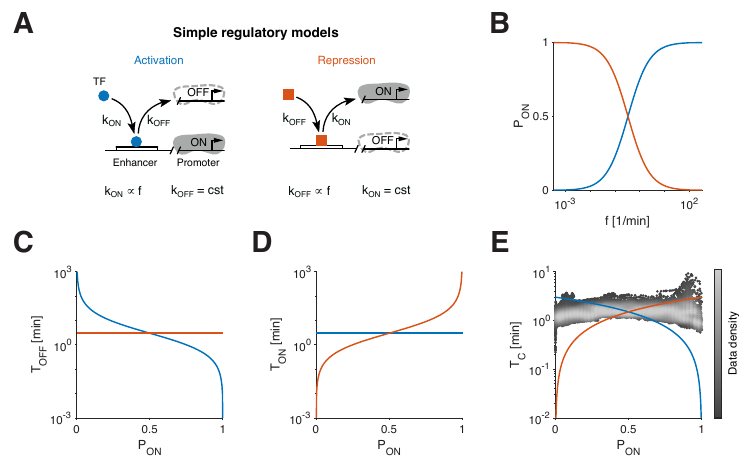}
\caption{{\bf Simple regulatory models with single TF binding site.}
{\bf (A)} Reaction scheme of the activation and repression models. These are simple two-state models in which the promoter switches between an active (ON) and inactive (OFF) state. One switching rate—either activation ($k_{\rm ON}$) or repression ($k_{\rm OFF}$), varies proportionally with the input transcription factor concentration $f$, while the other rate remains constant.
{\bf (B)} Induction functions, defined as the steadystate ON-probability $P_{\rm ON}=k_{\rm ON}/(k_{\rm ON}+k_{\rm OFF})$ as a function of $f$, are shown for the activation (blue) and repression model (orange).
{\bf (C)} Mean OFF-time $T_{\rm OFF}$, defined as $1/k_{\rm ON}$, plotted as function of $P_{\rm ON}$, color code as in B.
{\bf (D)} Mean ON-time $T_{\rm ON}$, defined as $1/k_{\rm OFF}$, plotted as function of $P_{\rm ON}$, color code as in B.
{\bf (E)} Switching correlation time, defined as $\tc = T_{\rm ON} T_{\rm OFF}/(T_{\rm ON}+T_{\rm OFF})$, plotted as a function of $P_{\rm ON}$. Neither the simple activation nor repression model reproduces the invariant $\tc$ observed in the data (gray scale indicates data density). Additionally, matching the experimental scale required setting $1/b = 3$ min, an unrealistically long TF residence time compared to typical values (1–10 s).}
\label{figS1}
\end{figure*}

\begin{figure*}[h!]
\centering
\includegraphics[scale=1.25]{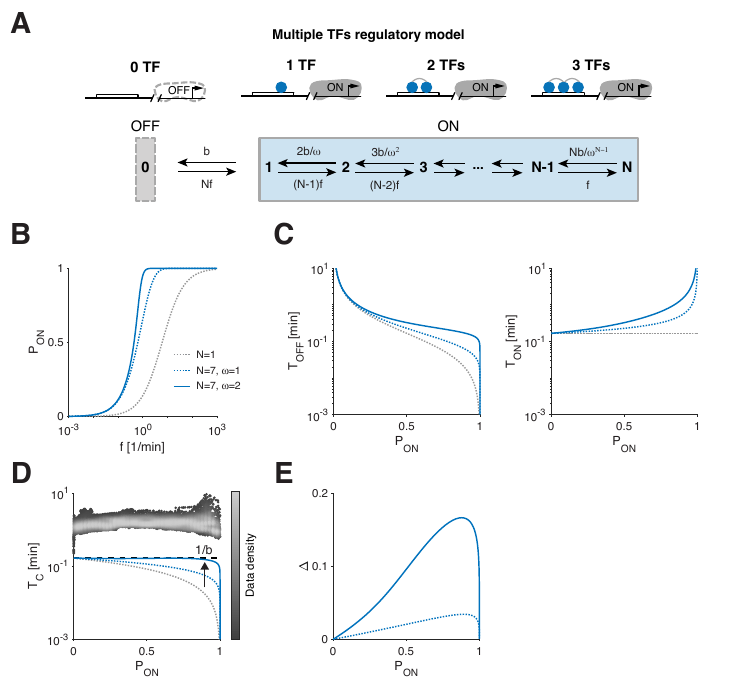}
\caption{{\bf Extending simple activation model to multiple transcription factor binding sites}
{\bf (A)} Reaction scheme for the activation model with $N$ transcription factor (TF) binding sites. The promoter is active (ON) when at least one TF is bound, and inactive (OFF) when none are bound. Up to $N$ TFs can bind simultaneously, each with rate $f$ (proportional to TF concentration), and unbind independently at rate $b/\omega^{n-1}$, where $b$ is the basal unbinding rate, $n \in {0, \dots, N}$ is the number of bound TFs, and  $\omega \geq 1$ is a cooperativity factor. Assuming identical binding sites (equal affinity), the system reduces by symmetry to a simple ladder model (bottom). The resulting system is an equilibrium model.
{\bf (B)} Induction functions, defined as the steady-state ON probability $P_{\rm ON}$ as a function of $f$, are shown for the simple activation model with $N=1$ binding site (see Fig.~\ref{figS1}), and for models with $N=7$ binding sites, both without cooperativity ($\omega=1$) and with cooperativity ($\omega=2$). Increasing either $N$ or $\omega$ sharpens the response, enhancing the system’s sensitivity to $f$.
{\bf (C)} Mean OFF-time $T_{\rm OFF}$ and ON-time $T_{\rm ON}$ as functions of $P_{\rm ON}$ for the same models as in (B). As $N$ and/or $\omega$ increase, the curves become increasingly antisymmetric around $P_{\rm ON} = 0.5$, particularly for $T_{\rm OFF}$.
{\bf (D)} Switching correlation time $\tc$ as a function of $P_{\rm ON}$ for the models in (C). As $N$ and/or $\omega$ increase, $\tc$ becomes flatter, approaching an invariant profile. However, it remains bounded by $1/b$, which is much lower than the average correlation time observed in the data ($\sim 1.5$ min, grayscale density) given realistic TF unbinding times ($1/b \leq 10$ s).
{\bf (E)} Distance $\Delta$ between multi-state and two-state models as function of $P_{\rm ON}$. As $N$ and/or $\omega$ increase, the distance grows, indicating that the multi-state $\tc$ is less accurately represented by a coarse-grained two-state model, even as flatness improves in (D). This suggests that simple multi-TF activation models are insufficient to explain the experimentally observed invariant $\tc$.}
\label{figS2}
\end{figure*}

\begin{figure*}[h!]
\centering
\includegraphics[scale=1.25]{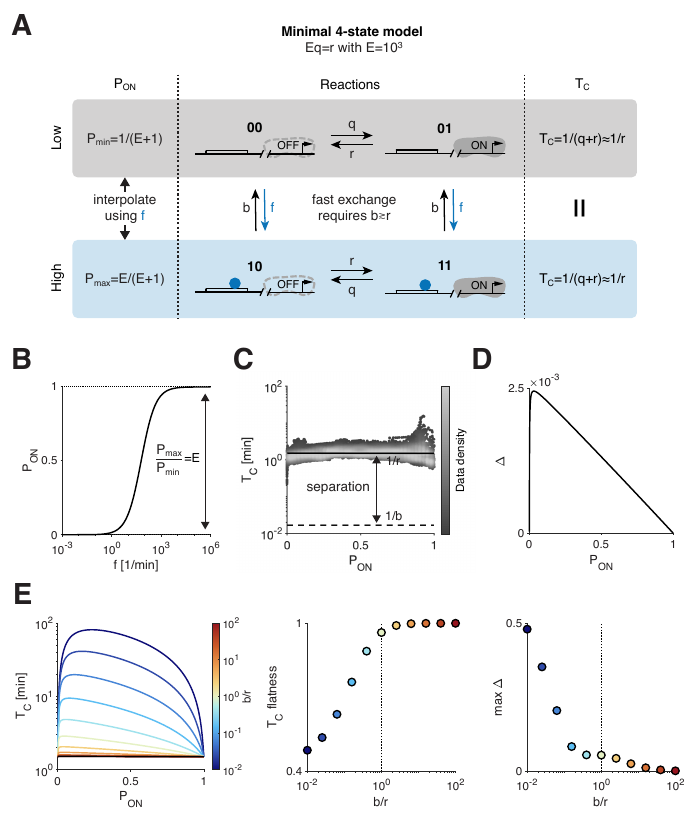}
\caption{{\bf Minimal mechanistic model achieving flat correlation time.}
{\bf (A)} Minimal reaction scheme for generating an invariant switching correlation time $\tc$ using only elementary reactions. The model combines two two-state systems with identical $\tc = 1/(q + r)$ but distinct ON probabilities: $P_{\rm min} = q/(q + r)$ and $P_{\rm max} = r/(q + r)$, achieved by swapping the forward ($q$) and backward ($r$) promoter switching rates. These systems are coupled through reversible transitions (with rates $f$ and $b$) representing TF binding and unbinding at the enhancer, enabling continuous control of $P_{\rm ON} \in [P_{\rm min}, P_{\rm max}]$ via the TF-dependent rate $f$. The resulting four-state model (states {\tt 00}, {\tt 10}, {\tt 01}, {\tt 11}) captures both TF binding and promoter activity, while preserving a constant $\tc$. Notably, because promoter switching rates are reversed at low and high activity, the model violates detailed balance and operates out of equilibrium.
{\bf (B)} Typical induction function, showing $P_{\rm ON}$ as a function of TF concentration $f$ for the minimal four-state model. By setting $P_{\rm min} = 1/(E + 1)$ and $P_{\rm max} = E/(E + 1)$, the model achieves a large expression dynamic range $E$, here set to $10^3$. This choice imposes $r=Eq$, implying $r \gg q$.
{\bf (C)} Switching correlation time $\tc$ as a function of $P_{\rm ON}$. The model naturally produces an invariant $\tc$, with its scale set approximately by $1/r$. Here, $1/b = 1$ s is used, and $1/r = 1.5$ min is chosen to match the experimental timescale (gray density). This results in a natural timescale separation, with $\tc = 1/r \gg 1/b$.
{\bf (D)} Distance $\Delta$ between the four-state and two-state models, plotted as a function of $P_{\rm ON}$. The distance remains small across the range, indicating that the four-state model effectively behaves as a two-state system. This supports the interpretation that the invariant $\tc$ observed in (C) reflects a genuine, coarse-grainable phenotype.
{\bf (E)} Effect of the $b/r$ ratio on the $\tc$ function and the effective two-state behavior of the four-state minimal model. (Left) $\tc$ versus $P_{\rm ON}$ for varying $b/r$ (color-coded) shows loss of $\tc$-invariance when $b \gtrsim r$. (Middle) Flatness decreases sharply as $b/r$ drops below 1, confirming the breakdown of invariance. (Right) The model becomes poorly coarse-grainable for $b/r < 1$, as indicated by a steep increase in the maximum $\Delta$. Together, these results show that maintaining $b \gg r$, i.e., a clear timescale separation, is essential for achieving both $\tc$-invariance and a meaningful two-state approximation.}
\label{figS3}
\end{figure*}

\begin{figure*}[h!]
\centering
\includegraphics[scale=1.25]{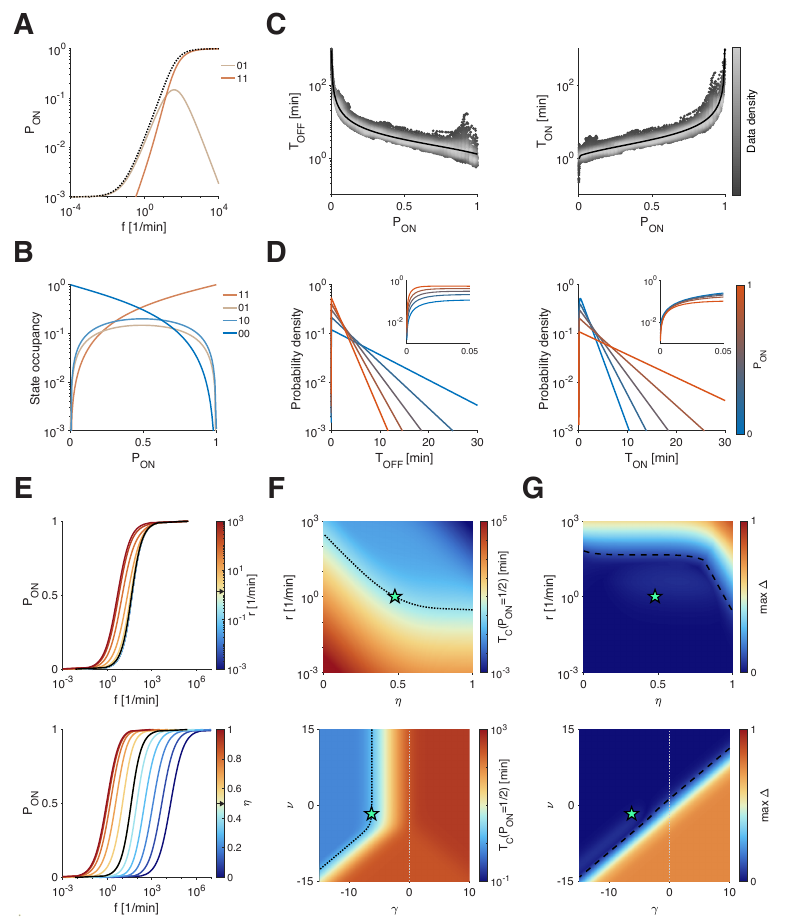}
\caption{{\bf Closer look at the four-state cycle model.}
{\bf (A)} Induction curve for the best-fit model shown in Fig.~\ref{fig2} (green star in B, D–E; black line in C). The overall ON probability $P_{\rm ON}$ is shown as a dotted line, with contributions from the active states {\tt 01} and {\tt 11} indicated in light and dark orange, respectively.
{\bf (B)} State occupancies of the four-state model ({\tt 00}, {\tt 10}, {\tt 01}, {\tt 11}) as a function of $P_{\rm ON}$ for the best-fit model (as in A). Active states are shown in orange; inactive states in shades of blue.
{\bf (C)} Mean OFF-time $T_{\rm OFF}$ and ON-time $T_{\rm ON}$ as functions of $P_{\rm ON}$ for the best-fit model (as in A), shown as black curves. The model captures the experimental trends well, as reflected by the data density in gray.
{\bf (D)} ON and OFF residence time distributions as $P_{\rm ON}$ varies for best-fit model (as in A). Consistent with an effective two-state model, these distributions are close to exponential, with deviations only at very short time scale (inset).
{\bf (E)} Induction curves as functions of $r$ and $\eta$ (color-coded). Black curves correspond to the best-fit model (as in A). Variations in $\eta$ have a more pronounced effect on the induction profile than changes in $r$.
{\bf (F)} Phase space of the switching correlation time $\tc$ at mid-expression ($P_{\rm ON} = 1/2$) as the control parameters ($r$, $\eta$, $\gamma$, $\nu$) vary. The green star marks the best-fit model (as in A), and the dotted line indicates the experimental value $\tc(P_{\rm ON} = 1/2) = 1.5$ min.
{\bf (G)} Phase space of the $\Delta$ distance between the four-state cycle model and the two-state model as the control parameters ($r$, $\eta$, $\gamma$, $\nu$) vary. The green star indicates the best-fit model (as in E), and the dashed line marks the threshold $\Delta = 0.1$, below which models are considered well coarse-grainable. This threshold also defines the shaded region shown in Fig.~\ref{fig2}E.}
\label{figS4}
\end{figure*}

\begin{figure*}[h!]
\centering
\includegraphics[scale=1.25]{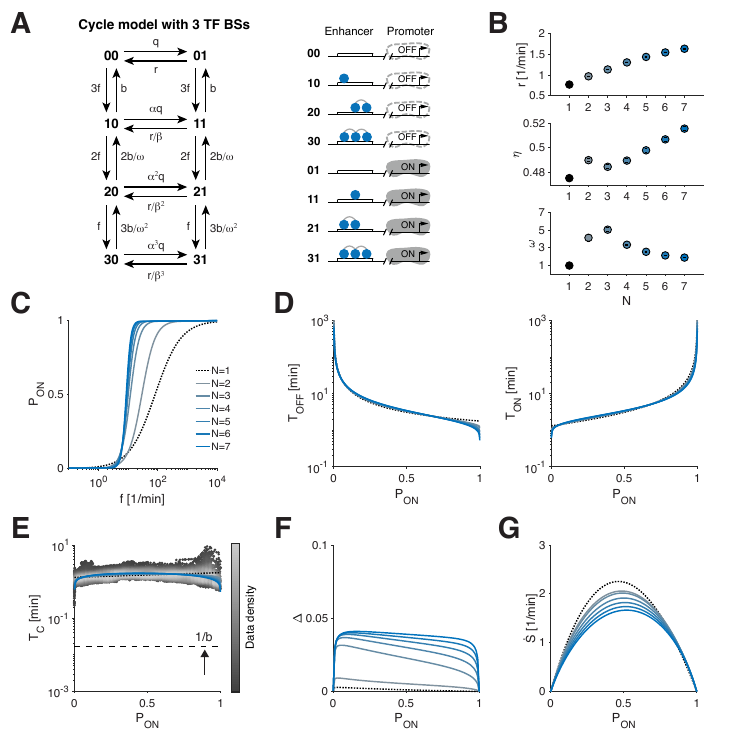}
\caption{{\bf Extending the cycle model to multiple transcription factor binding sites}
{\bf (A)} Extension of the cycle model to include $N$ cooperative TF binding sites at the enhancer (illustrated for $N = 3$). TFs bind at rate $f$ and unbind at rate $b/\omega^{n-1}$, where $n$ is the number of bound TFs and $\omega \geq 1$ is a cooperativity factor. Promoter activation and deactivation occur at rates $\alpha^n q$ and $r/\beta^n$, respectively, allowing the system to benefit from multiple bound TFs to enhance activation and stabilize the ON state. Assuming identical binding sites (equal affinity), the reaction network reduces by symmetry to a chain of four-state cycle models (left), with individual states labeled as in the original model (right). Parameters are constrained to span a fixed range of $P_{\rm ON}$, with three free parameters: $(r, \eta, \omega)$. The parameter $\eta$ controls $\alpha$ and $\beta$ as in the original model: $\alpha = E^{2\eta/N}$ and $\beta = E^{2(1-\eta)/N}$, where $E = P_{\rm max}/P_{\rm min}$ defines the dynamic range.
{\bf (B)} Best-fit parameters $r$, $\eta$, and $\omega$ as a function of the number of binding sites $N$, obtained using the same inference approach as for the original model. Estimates represent the median of the posterior distribution, with error bars indicating the 95\% credible interval.
{\bf (C)} Induction curves for the best-fit models in (B), shown as a function of the number of binding sites $N$ (color-coded as in B). As $N$ increases, the curves become steeper, reflecting greater sensitivity to the input TF concentration $f$.
{\bf (D)} Mean OFF-time $T_{\rm OFF}$ and ON-time $T_{\rm ON}$ as functions of $P_{\rm ON}$ for the best-fit models in (B). Despite increasing $N$, these functions remain nearly unchanged and approximately antisymmetric.
{\bf (E)} Switching correlation time $\tc$ as a function of $P_{\rm ON}$ for the best-fit models in (B). The extended models maintain an approximately invariant $\tc$ across $P_{\rm ON}$, even as $N$ increases. Time scale separation is preserved, with $1/b = 1$ s and $\avg{\tc} \sim 1.5$ min.
{\bf (F)} Distance $\Delta$ between the extended cycle models (best-fit, as in B) and corresponding two-state models, plotted as a function of $P_{\rm ON}$. Despite the increasing number of microstates with higher $N$, $\Delta$ remains small, indicating that the extended models retain effective two-state behavior.
{\bf (G)} Entropy production rate $\dot{S}$ as a function of $P_{\rm ON}$ for the best-fit models in (B). All extended models operate out of equilibrium. Interestingly, as $N$ increases, dissipation slightly decreases.}
\label{figS5}
\end{figure*}

\begin{figure*}[h!]
\centering
\includegraphics[scale=1.25]{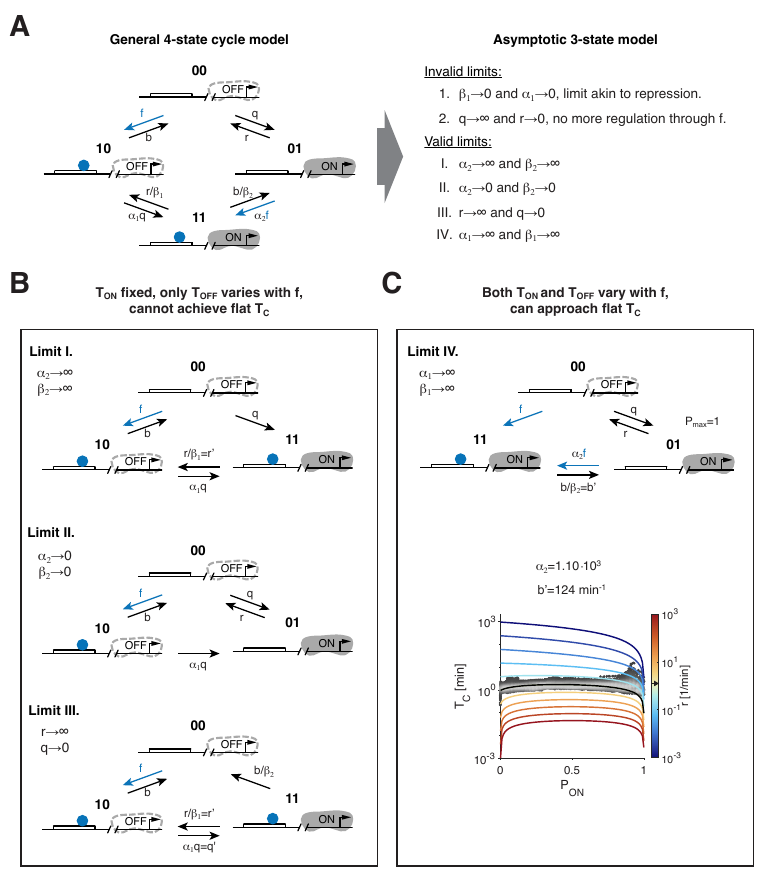}
\caption{{\bf Three-state model cannot generate invariant correlation time.}
{\bf (A)} Three-state models can be derived as asymptotic limits of the mechanistic four-state cycle model (left). By taking specific pairs of elementary rates to zero and/or infinity, four valid three-state limits emerge. However, these models are not fully mechanistic, as one transition in each involves an infinitely fast intermediate step—making the reaction effectively non-elementary.
{\bf (B)} In most three-state models (limits I–III from A), only a single reaction depends on the TF concentration $f$, corresponding to TF binding. As a result, only $T_{\rm OFF}$ varies with $f$, while $T_{\rm ON}$ remains constant. This is insufficient to generate an invariant switching correlation time $\tc$, which requires anti-correlated variation in both $T_{\rm OFF}$ and $T_{\rm ON}$.
{\bf (C)} Only one limit (limit IV in A, with $\alpha_1, \beta_1 \rightarrow \infty$) yields a three-state model in which two reactions depend on $f$, allowing both $T_{\rm OFF}$ and $T_{\rm ON}$ to vary. While this configuration can approximate an invariant $\tc$ and fit the data reasonably well, it still lacks a true mechanistic basis: it hides an infinitely fast intermediate step, and the reverse transition (${\tt 11} \rightarrow {\tt 00}$) is absent. Notably, adding this missing reverse reaction worsens $\tc$ flatness.}
\label{figS6}
\end{figure*}

\begin{figure*}[h!]
\centering
\includegraphics[scale=1.25]{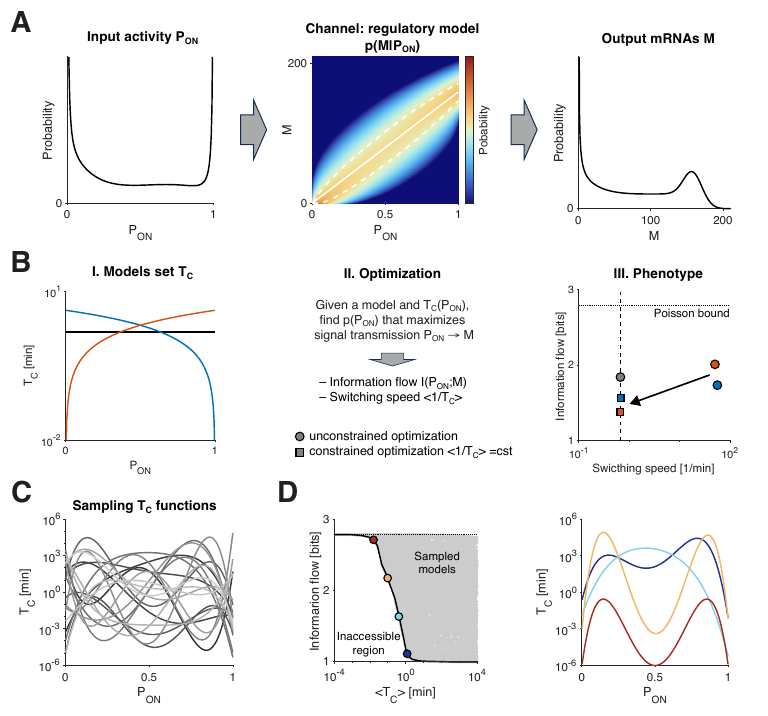}
\caption{{\bf Optimizing information flow in gene regulatory models.}
{\bf (A)} Schematic of information flow in gene regulation. Regulatory models are treated as input-output channels (left to right), where the transcription factor concentration $f$ is the true input and the steady-state mRNA copy number $M$ is the output. Promoter activity $P_{\rm ON}(f)$ serves as a convenient proxy for the input, as it is a monotonic function of $f$. Each regulatory model is coupled to an mRNA birth-death process with transcription initiation rate $K$ and mRNA lifetime $T_{\rm M}$, allowing computation of the channel conditional distribution $p(M|P_{\rm ON})$, shown in the center. Biologically realistic parameters are used, based on estimates for Drosophila gap genes ($K = 6$, $T_{\rm M} = 20$ min).
{\bf (B)} Overview of the optimization procedure. (I) Each model is defined by its switching correlation time function $\tc(P_{\rm ON})$, which fully characterizes a two-state system. (II) For a given model, we compute the information flow $\Phi$ by optimizing the mutual information $I(P_{\rm ON}; M)$ over input distributions $p(P_{\rm ON})$ using the Arimoto-Blahut algorithm. We assume no input constraint so that information flow is limited solely by the model’s dynamics. We then compute the mean switching speed $V=\avg{1/\tc}$, which quantifies the dynamical efficiency of the system. (III) Models are then classified in terms of their phenotype in the $(\Phi, V)$ space, allowing us to compare the trade-off between information transmission and switching speed across regulatory strategies.
{\bf (C)} Random sampling of switching correlation time functions $\tc(P_{\rm ON})$ using a degree-four orthogonal polynomial basis. Sampling is performed in log-space over a broad dynamic range ($\tc \in [10^{-6}, 10^6]$), allowing exploration of the full space of generic two-state models, which are fully defined by their $\tc$ profile. A total of $5.5 \times 10^4$ random models were generated, each yielding corresponding values of information flow $\Phi$ and switching speed $V$. This approach reveals the fundamental limits and trade-offs between information transmission and dynamic responsiveness in regulatory systems.
{\bf (D)} (Left) Information flow $\Phi$ as a function of mean switching time $\avg{\tc}$, computed across the space of randomly sampled two-state models from (C). This view reveals the worst-performing models as a lower Pareto front (black line), corresponding to those with minimal information at a given switching timescale. (Right) $\tc$ as a function of $P_{\rm ON}$ for four representative models along the Pareto front (left). These curves are far from flat, illustrating that unlike switching speed $V$ (used in Fig.~\ref{fig3}A to identify optimal models), the average switching time $\avg{\tc}$ fails to capture the presence of invariant $\tc$.}
\label{figS7}
\end{figure*}

\begin{figure*}[h!]
\centering
\includegraphics[scale=1.25]{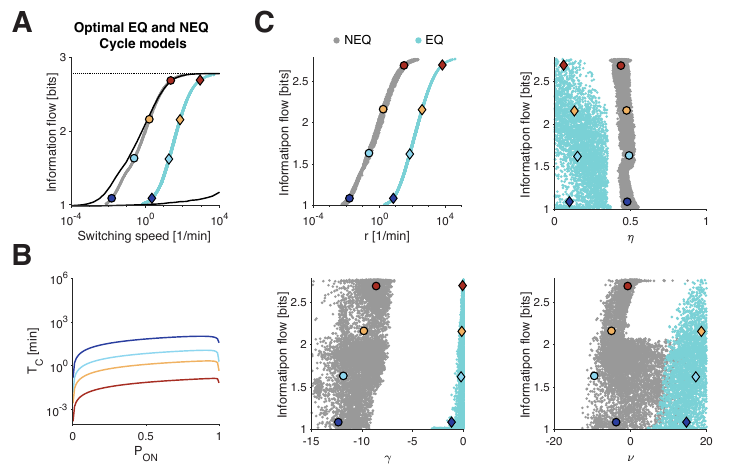}
\caption{{\bf Optimal cycle models.}
{\bf (A)} Optimal non-equilibrium (NEQ, gray) and equilibrium (EQ, cyan) cycle models in the $(\Phi, V)$ space. NEQ models clearly outperform EQ models, achieving higher information flow $\Phi$ at the same switching speed $V$, or lower $V$ at the same $\Phi$. EQ models are defined here as those with very low entropy production ($\dot{S} < 10^{-4}$), a slightly looser criterion than the strict equilibrium condition $\gamma = 0$. The best-performing NEQ models closely approach the Pareto front of generic two-state systems (see Fig.~\ref{fig3}A–B), illustrating the functional advantage of non-equilibrium regulation.
{\bf (B)} $\tc$ curves as a function of $P_{\rm ON}$ for the optimal EQ models (colored diamonds from A). The resulting functions are far from flat, reflecting the inherent inability of equilibrium models to generate invariant switching correlation times.
{\bf (C)} Information flow $\Phi$ as a function of the model parameters for the optimal NEQ and EQ models in (A). The two classes occupy distinct regions in parameter space, underscoring the structural differences between EQ and NEQ strategies.}
\label{figS8}
\end{figure*}

\begin{figure*}[h!]
\centering
\includegraphics[scale=1.25]{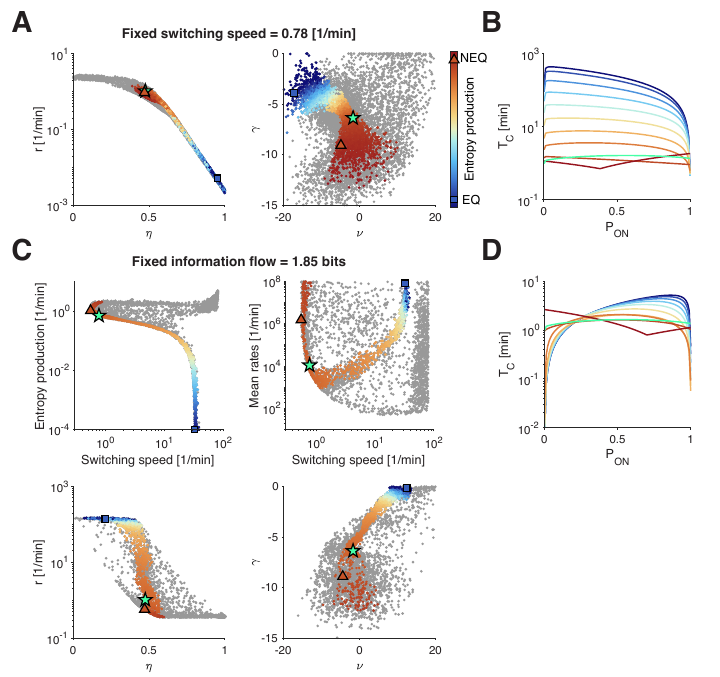}
\caption{{\bf Navigating trade-offs around best-fit model.}
{\bf (A)} Parameters corresponding to a continuous path from equilibrium (EQ) to non-equilibrium (NEQ) models that maximize information flow at fixed switching speed $V = 0.78$ min$^{-1}$, matching the value of the best-fit model (see Fig.~\ref{fig3}D). Color-coded dots represent models along this path as dissipation varies, while gray dots indicate sampled models outside the path. The green star marks the location of the best-fit model.
{\bf (B)} Typical $\tc$ functions for models along the EQ-to-NEQ path at constant switching speed (see A). Colors correspond to dissipation levels shown in (A), and the green curve represents the best-fit model. As dissipation decreases, the $\tc$ functions deviate substantially from being a constant. Near equilibrium, the fixed switching speed $V$ is maintained primarily through low $\tc$ values near $P_{\rm max}$.
{\bf (C)} Section of the regulatory phenotype manifold (top) and corresponding parameter space (bottom), obtained by fixing the information flow to 1.85 bits, matching the value of the best-fit model (see Fig.~\ref{fig3}C). Color-coded dots represent a continuous path from equilibrium to non-equilibrium models that minimizes switching speed at fixed information flow, with color indicating dissipation. Gray dots correspond to sampled models outside this path. The green star marks the best-fit model.
{\bf (D)} Typical $\tc$ functions for models along the EQ-to-NEQ path at constant information flow (see C). As dissipation varies, the deformation of the $\tc$ curves is less pronounced than in the fixed switching speed case (B), since maintaining constant information flow enforces an approximately fixed average $\tc$. However, in the near-equilibrium regime, portions of the curves can reach very low $\tc$ values, as switching speed $V$ is not constrained in this setting.}
\label{figS9}
\end{figure*}

\begin{figure*}[h!]
\centering
\includegraphics[scale=1.25]{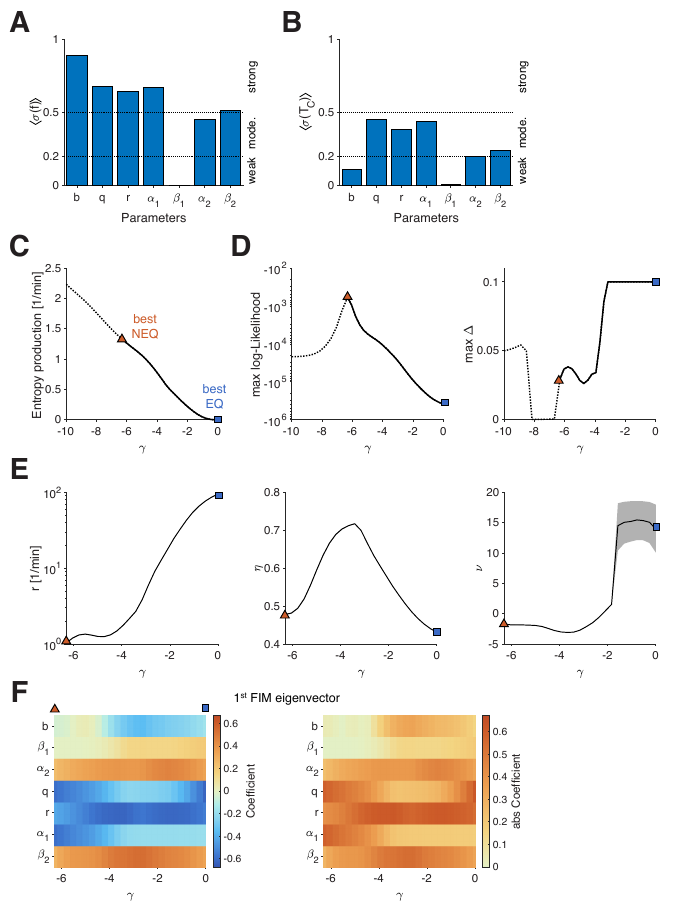}
\caption{{\bf Model perturbations and comparing best-fit models.}
{\bf (A-B)} Sensitivity of the induction curves $f$ (A) and switching correlation time functions $\tc$ (B) to 10-fold perturbations of individual parameters. Perturbations are applied as Gaussian noise in log-space. For each parameter, the response $\sigma$ is defined as the root mean squared displacement between the reference curve (either $f$ or $\tc$) with best-fit parameters and the perturbed curve, computed in log-space and averaged over $P_{\rm ON}$. The overall size effect $\langle \sigma \rangle$ is then calculated as the mean response across multiple realizations of log-normal noise applied independently to each parameter. A summary of size effects is shown in Fig.~\ref{fig4}B.
{\bf (C-D)} Varying $\gamma$ to control the entropy production of best-fit models. Resulting $\tc$ functions are shown in Fig. \ref{fig4}C and corresponding parameters in E. (C) Entropy production decreases approximately linearly as $\gamma$ increases. (D) Maximum log-Likelihood is maximal for $\gamma=-6.32$ corresponding to the overall best-fit NEQ model (see also Fig. \ref{fig2}B), while the best EQ model at $\gamma=0$ clearly provides a worse fit to data. We enforced that all best-fit model behaves as 2-state model ($\max \Delta\leq0.1$). 
{\bf (E)} Parameters of best-fit models displayed in Fig. \ref{fig4}C, as $\gamma$ varies, thus adjusting the entropy production (see C).
{\bf (F)} First eigenvector of the Fisher Information Matrix (FIM) associated with the dominant eigenvalue as $\gamma$ varies. The vector at the two $\gamma$ end-points are displayed in Fig. \ref{fig4}E.}
\label{figS10}
\end{figure*}

%TC:endignore
\end{document}